\documentclass{article}

\usepackage[preprint, nonatbib]{neurips_2022}

\usepackage[utf8]{inputenc} % allow utf-8 input
\usepackage[T1]{fontenc}    % use 8-bit T1 fonts
\usepackage{hyperref}       % hyperlinks
\usepackage{url}            % simple URL typesetting
\usepackage{booktabs}       % professional-quality tables
\usepackage{amsfonts}       % blackboard math symbols
\usepackage{nicefrac}       % compact symbols for 1/2, etc.
\usepackage{microtype}      % microtypography
\usepackage{color,soul}    
\usepackage{graphicx}       % to include images
\usepackage{verbatim}       % block-level comments
\usepackage[table]{xcolor}
\usepackage{csvsimple}
\usepackage{siunitx}
\usepackage{booktabs}
\usepackage{tabularx}
\usepackage{float}
\usepackage{makecell}
\usepackage{geometry}
\usepackage{import}
\usepackage{pgfplots}
\usepackage{subcaption}
\pgfplotsset{compat=1.17}
\usepackage{enumitem}

\title{Training self-supervised peptide sequence models on artificially chopped proteins}

\author{%
  Gil Sadeh\thanks{These authors contributed equally} \\
  Amazon Devices\\
  \texttt{gilsadeh@amazon.com} \\
   \And
   Zichen Wang\footnotemark[1] \\
   Amazon Machine Learning Solutions Lab \\
   \texttt{zichewan@amazon.com} \\
   \And
   Jasleen Grewal\footnotemark[1] \\
   AWS Professional Services \\
   \texttt{gjaslee@amazon.com} \\
   \And
   Huzefa Rangwala \\
   Amazon Machine Learning Solutions Lab \\
   \texttt{rhuzefa@amazon.com} \\
   \And
   Layne Price\thanks{Corresponding author} \\
   Amazon Devices \\
   \texttt{prilayne@amazon.com} \\
}

\begin{document}

\maketitle

\begin{abstract}

Representation learning for proteins has primarily focused on  the global understanding of protein sequences regardless of their length. However, shorter proteins (known as peptides) take on distinct structures and functions compared to their longer counterparts. Unfortunately, there are not as many naturally occurring peptides available to be sequenced and therefore less peptide-specific data to train with.
In this paper, we propose a new peptide data augmentation scheme, where we  train peptide language models on artificially constructed peptides that are small contiguous subsets of longer, wild-type proteins; we refer to the training peptides as “chopped proteins”. We evaluate the representation potential of models trained with chopped proteins versus natural peptides and find that training language models with chopped proteins results in more generalized embeddings for short protein sequences. These peptide-specific models also retain information about the original protein they were derived from better than language models trained on full-length proteins. 
We compare masked language model training objectives to three novel peptide-specific training objectives: next-peptide prediction, contrastive peptide selection and evolution-weighted MLM.
% We demonstrate state-of-the-art results on the common TAPE stability benchmark, which consists of peptides of length less than 50 amino acids.  
We  demonstrate improved zero-shot learning performance  for a  deep mutational scan  peptides benchmark.

\end{abstract}

\section{Introduction}

Proteins are integral to all cellular functions in living organisms. 
%Long, organized concatenation of basic molecular building blocks (amino acids) referred by primary sequences form complex structures that can perform specialized functions like maintaining cell structure or driving biochemical reactions. 
Proteins of length less than 50 amino acids, called peptides, can have different functional properties from larger proteins partly due to physical characteristics (shorter length and simpler three-dimensional folded structures) and partly due to their distinct biological or therapeutic roles\cite{FOSGERAU2015122, muttenthaler2021trends}. 
Naturally occurring peptides typically act as chemical messengers within and outside cells by interacting with other molecules like cellular receptors or antigens~\cite{apostolopoulos2021global}.
They also play key roles as interaction partners and messengers in the immune system of multicellular organisms \cite{morimoto2018therapeutic}. 
%Due to this, peptides serve as chemical messengers mediating cellular function by interacting with other molecules like cellular receptors or antigens .
Given their functional properties and ease of modification using laboratory techniques, peptide drug development has emerged as a promising area of therapeutic research and development \cite{wang2022therapeutic}.

While there has been several prior works in self-supervised protein models \cite{esm, prottrans, ferruz2022deep, heinzinger2019modeling, alley2019unified, bepler2019learning, he2021pre, madani2020progen, detlefsen2022learning, rao2020transformer}, unfortunately the ability of these models to represent peptide sequences is understudied, as there are relatively few known naturally occurring peptides to experiment with.
Few works~\cite{cheng2021pepformer, serrano2020deepmspeptide} have focused on peptide sequences, training transformer and convolutional models to predict peptide detectability by mass spectrometry. However, these methods involve supervised training over mass-spec data, and do not explore leveraging 
large-scale proteomic data with self-supervised learning.

% : for example, Uniprot/UniParc\cite{uniprot2019uniprot} has >500 million proteins of length >50 and 15 million (~3\%) peptides of length <50. 
% \gs{removed example with numbers as this is not the actual numbers in our uniref version (we use an older version to be compatible with ESM) }

In this work we train self-supervised learning models for peptides inspired from natural language modeling with the input representation being the primary sequence. 
%language-style models on peptides, as represented by their primary sequence of amino acids. 
We propose a new peptide-specific data augmentation framework, which we call ``chopped proteins.'' %Additionally, we would like the peptide models to generalize to mutated peptides that are close, but not identical to their wild-type counterparts, as well as designed peptides. \jg{not sure I follow this last sentence - do we split it into two, rationalizing along the way why we want them to be able to generalize to mutated variants distinct from wild-type, and what we mean by generalizing to designed peptides?} \jg{DELETE THIS COMMENTED SET OF LINES IF THE NEXT SENTENCE IS ACCEPTABLE}
%Additionally, we would like the peptide models to generalize to novel variants with distinct biology from wild-type counterparts, such as therapeutic formulations or laboratory designed peptides. 
 Our data augmentation strategy involves simulating the primary sequence of novel peptides by randomly sampling contiguous subsets of longer protein primary sequences available in common databases, such as UniRef\cite{suzek2007uniref}. This sub-sampling scheme is inspired by intracellular proteasomal cleavage mechanisms\cite{tanaka2009proteasome}, where proteins are degraded by being chopped into smaller peptides by protease enzymes. We train and evaluate the peptide models on chopped proteins using different training objectives. We use standard masked language modeling (MLM)~\cite{devlin2018bert} objective, and explore additional alternatives, exploiting pairwise peptide relationships and prior evolutionary information.

\section{Methods}
\label{sec:methods}

\paragraph{Baseline model, ESM1b.---}
ESM1b~\cite{esm} is a protein transformer model, composed of $33$  layers with embedding dimension $d=1280$. It was trained over ~30M cluster representative amino-acid sequences (i.e. primary structures), from the UniRef50~\cite{suzek2007uniref} dataset, using BERT-like MLM self-supervised training.
ESM1b has previously achieved state-of-the-art performance, in comparison to competitive sequence based models, in mutational effect, secondary structure, and long-range contact prediction tasks. 
In addition, the authors have released the pretrained model weights \footnote{ \url{https://github.com/facebookresearch/esm}}.
Hence, we have chosen to use this model as our baseline, while utilizing the same architecture and initialization from the provided pretrained weights for our peptide-based training.

%\begin{figure}[t]
%    \centering
%    \includegraphics[width=0.65\textwidth]{sections/methodfigures/MLM.png}
%    \caption{\textbf{MLM self-supervised pretraining procedure for training on chopped proteins}. }
%    \label{fig:pmlm}
%\end{figure}

\paragraph{Chopped proteins from UniRef50.---}
In order to apply large scale self-supervised training for short length peptide sequences, we utilize available large protein data, and randomly "chop" small contiguous subsets from the given long protein sequences, as illustrated in Figure~\ref{fig:pmlm}. 
When "chopping", peptide sequence length is uniformly sampled from 8 to 50. As in ESM1b, we also utilize the UniRef50~\cite{suzek2007uniref} dataset. 
UniRef database provides clustered sets of sequences (from UniProt and UniParc) to obtain complete coverage of sequence space at several resolutions. UniRef50 consists of clustered sequences at 50\% sequence identity level while hiding redundant sequences. Like ESM1b, we only utilize cluster representative sequences, which contain the most biological information. This dataset, also known as UniRef50-sparse (UR50-S), contains 30M cluster representative protein sequences. We use the test set provided by ESM1b and randomly split the remaining sequences between training (90\%) and validation (10\%) sets.

\paragraph{Masked Language Modeling (MLM).---}
This objective, originally proposed in BERT~\cite{devlin2018bert}, is commonly used for learning language and protein representations~\cite{esm, prottrans, heinzinger2019modeling, alley2019unified}. For each sequence x, we sample a set of indices $M$ to mask, with probability p=0.15, replacing the true token at each index $i$ of the input sequence with either (1) the <MASK> token (p=0.8); (2) a random amino acid token (p=0.1); or (3) the unchanged $i^\mathrm{th}$ token (p=0.1). For each index $i \in M$, we independently minimize the negative log likelihood of the true amino acid $x_i$ given the masked input sequence $x_{/M}$ as context: $\mathcal{L}_{MLM} = - \mathbb{E}_M[\sum_{i \in M} (\log{p(\hat{x}_i=x_i|x_{/M})}]$
, where $\log{p(\hat{x}_i=x_i|x_{/M})}$ is the logit corresponding to the true $i^\mathrm{th}$ token. Figure~\ref{fig:pmlm} illustrates applying MLM on chopped proteins.

\paragraph{Alternative training objectives.---}
We additionally explore several alternative training objectives. We consider complementing the MLM objective with an additional, auxiliary, sequence-level, objective for learning pairwise peptide relationship. We propose the 'Next-Peptide-Prediction' (NPP) objective, which is equivalent to BERT's 'Next-Sentence-Prediction' objective, and an additional contrastive extension of it (C-NPP). In addition, we propose the BLOSUM MLM (BMLM) objective, which incorporates evolutional information utilizing the Blosum62 substitution matrix\cite{henikoff1992amino} into the MLM training. These objectives are described in detail in appendix section~\ref{sec:appendix-training-objectives}.

\section{Results}

%\subsection{Experimental Protocol}

Unless stated otherwise, all our peptide models use the ESM1b backbone architecture and initialization, and were trained similarly, for 25 epochs over protein sequences, saving the best performing checkpoint over the validation set. We used Adam optimizer, with 0.0001 base learning rate, inverse square-root decay schedule, 10K warmup steps and an effective batch size of 8192 sequences.

\ExplSyntaxOn
\NewDocumentCommand { \getminece } { }
  {
    \clist_gset:Nx \g_tmpa_clist {\esmb, \mlm, \mlmnsp, \contrastive, \bmlm}
    \clist_sort:Nn \g_tmpa_clist
      {
        \fp_compare:nNnTF {##1} > {##2}
          { \sort_return_swapped: }
          { \sort_return_same: }
      }
    \tl_gset:Nx \g_tmpa_tl { \clist_item:Nn \g_tmpa_clist {1} }
  }
\NewDocumentCommand { \mynum } { m }
  {
    \fp_compare:nNnTF { #1 } = { \g_tmpa_tl }
      { \cellcolor{blue!20} \num{#1} }
      { \num{#1} }
  }
\ExplSyntaxOff

\def\mywidth{1.9cm}

\noindent\begin{table*}[t]
  \sisetup{round-mode=places, round-precision=5}
  \centering
  \caption{\textbf{Exponentiated Cross Entropy (ECE) results for sequence models using an MLM or derivative objective.} Lower value is better, lowest value for each trained model is shown in bold. }
%   \csvreader[tabular=lP{\mywidth}P{\mywidth}P{\mywidth}P{\mywidth}P{\mywidth}P{\mywidth}P{\mywidth}P{\mywidth},
%             table head=\toprule \thead{ Evaluation \\set} & \thead{ESM-1b} & \thead{Pept.\\MLM} & \thead{Pept.\\NPP} & \thead{Pept.\\C-NPP} & \thead{Pept.\\BMLM} \\ \midrule,
%             head to column names,
%             before line=\getminece,
%             %late after line = \\\hline,
%             late after last line=\\\bottomrule]
%             {sections/resultfiles/ECE_results_no_std.csv}%
%             {}%
%             {\csvcoli & \mynum{\esmb} & \mynum{\mlm} &
%             \mynum{\mlmnsp} & \mynum{\contrastive} & \mynum{\bmlm} 
%             }
    \begin{tabular}{p{\mywidth}p{\mywidth}p{\mywidth}p{\mywidth}p{\mywidth}p{\mywidth}p{\mywidth}p{\mywidth}}
    \hline
    \thead{Evaluation\\set} & \thead{ESM1b} & \thead{Pept.\\MLM} & \thead{Pept.\\NPP} & \thead{Pept.\\C-NPP} & \thead{Pept.\\BMLM} \\ \hline
    peptides & 6.6144$\pm$0.0073 & 6.3136$\pm$0.0063 & \textbf{6.3131}$\pm$0.0074 & 6.3386$\pm$0.0048 & 6.5331$\pm$0.0071 \\
    proteins & 4.8184$\pm$0.0087 & 4.4207$\pm$0.0095 & \textbf{4.3937}$\pm$0.0061 & 4.5204$\pm$0.0020 & 4.5523$\pm$0.0049 \\ \hline
    \end{tabular}
  \label{tab:eceresults}
\end{table*}

\paragraph{Peptide models have improved generalization to hold-out peptides and proteins---}  
% All of our supervised evaluations protocols follow the linear probing procedure instead of fine-tuning the language models. This ensures a fair evaluation of the learned representations without feature distortion by out-of-distribution data \cite{kumar2022fine}.   
The most straightforward evaluation for trained sequence models is to assess their performance on hold-out sequences using the exponentiated cross entropy (ECE) metric. ECE is the exponential of the model’s MLM loss averaged per token ($2^{\mathcal{L}_{MLM}}$). The ECE metric is analogous to perplexity commonly used for evaluating language models.
% It ranges from 1 for an oracle to vocabulary size (number of possible amino-acids / special tokens) for a uniform random prediction. 
% \gs{i changed the direct reference to vocabulary size being 25 as it's inaccurate, since we used tied embedding to output distibution also covers special tokens is actually 33, even though practically some of those tokens are unused.}
We report ECEs for all models over two subsets of UR50-S test set composed of peptides (short protein sequences) and proteins (of all lengths), respectively (Table~\ref{tab:eceresults}). The peptide models achieved markedly better ECEs than ESM1b on both evaluation sets.

\paragraph{Peptide models predict amino acids flanking a peptide sequence---}

\begin{figure}[t]
\caption{\textbf{Peptide context prediction accuracy over different peptide length bins.} 
% We plot accuracy and ECE, zooming in on differences between peptide transformer models in ECE
Average accuracy across 4 masked out context residues, beyond the peptide's edges, is shown.
} 
% \lp{can we make this figure two panels? it's a bit too big... is there something else we can show here?}
\centering
\begin{tikzpicture}[scale=0.7]
\begin{axis}[
    width  = 0.7\linewidth,
    height=6cm,
    xlabel={Peptide length},
    x label style={anchor=north, below=2mm},
    ylabel={Accuracy},
    x tick label style={rotate=45},
	symbolic x coords={8-16, 16-32, 32-64, 64-128, 128-256, 256-512},
    major x tick style = transparent,
	xtick=data,
    legend pos=north west,
    legend style={nodes={scale=0.7, transform shape}},
    ymajorgrids=true,
    grid style=dashed,
    ymin=0.07,
    ymax=0.23,
]

\addplot[
    color=blue,
    mark=square,
    ]
    coordinates {
    (8-16, 0.08603)(16-32,0.10473)(32-64,0.12392)(64-128,0.14395)(128-256,0.15887)(256-512,0.16765) 
    };

\addplot[
    color=green,
    mark=star,
    ]
    coordinates {
    (8-16, 0.12414)(16-32,0.14441)(32-64,0.16634)(64-128,0.1851)(128-256,0.19902)(256-512,0.20989) 
    };
\addplot[
    color=red,
    mark=triangle,
    ]
    coordinates {
    (8-16, 0.11778)(16-32,0.14123)(32-64,0.1639)(64-128,0.18509)(128-256,0.19886)(256-512,0.20988) 
    };
\addplot[
    color=brown,
    mark=*,
    ]
    coordinates {
    (8-16, 0.12404)(16-32,0.1438)(32-64,0.16535)(64-128,0.18253)(128-256,0.19405)(256-512,0.20267) 
    };
\addplot[
    color=gray,
    mark=diamond,
    ]
    coordinates {
    (8-16, 0.1217)(16-32,0.14173)(32-64,0.16456)(64-128,0.18459)(128-256,0.19954)(256-512,0.21223) 
    };
\legend{ESM-1b large, Pept. MLM, Pept. NPP, Pept. C-NPP, Pept. BMLM}    

\end{axis}
\end{tikzpicture}
\label{fig:peptidecontext}
\end{figure}
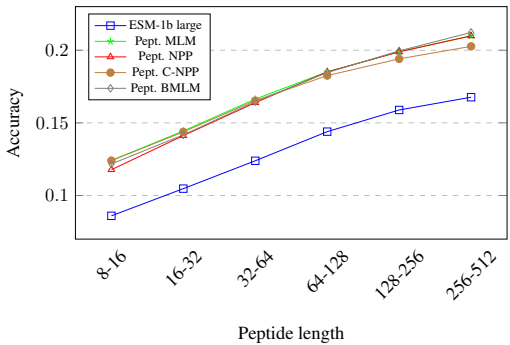

% \lp{An orphaned sentence from methods to incorporate}:
%     The objective of this task is to identify the predicted context token with the ground truth being the original masked out peptide. For this task we run a forward pass without any further fine tuning.   The evaluation is performed on a  a subset of the UR50-S test set, which was acquired by filtering out proteins shorter than 400 amino-acids. The filtering is done to ensure a fair comparison across different peptide length bins, by sampling peptides, of the various lengths, from the same set of protein sequences. The final evaluation is performed over peptides sampled from the filtered set of 706,240 proteins (one sampled peptide per protein).
    
% We test model's ability to predict which additional amino-acids were located on each side of the chopped peptide in the original protein they were derived from. 
%
By padding peptide sequences with mask tokens, and attempting to recover the correct amino-acids found in corresponding positions in the origin protein sequences, we evaluate model's ability to predict peptide context (also referred to as peptide's flanking regions). Despite similarity to MLM, this task is more challenging due to the selection of consecutive regions on the edges as the masked tokens. In addition, this task has biological significance as a peptide's context plays an important role in proteasomal cleavage~\cite{tanaka2009proteasome}. The evaluation was performed on randomly chopped sequences from a subset of long UR50-S test sequences on various sequence length bins, as shown in Figure~\ref{fig:peptidecontext}.
% The BloSim model is excluded from this evaluation as it was not trained with a MLM or equivalent loss component.
The results suggest that our peptide transformer models significantly outperform the baseline ESM1b protein model across short and long sequences. However, no statistically significant performance difference was found between the various training objectives.
Further, it is important to note that context prediction accuracy consistently improves as sequence length increases, with all models, despite peptide transformer models being only fine-tuned on short peptides, thus suggesting these models may preserve information learned from full proteins when fine-tuned on short peptides.

\paragraph{Functional analysis with zero-shot mutation effect prediction.---}
Evaluating the effects mutant protein sequence over its Wild-Type (WT) counterpart is a fundamental problem for understanding and designing proteins. Previous study \cite{meier2021language} found that pretrained protein language model can be used to score mutational effects without any training (zero-shot transfer).
We took 36 previously published deep mutational scans (DMS)~\cite{riesselman2018deep}, each of which experimentally quantifies a set mutant sequences over the WT protein. The goal of this task is to regress the mutant effects over its WT with zero-shot transfer. The WT marginal scoring scheme previously used in the ESM1v publication~\cite{meier2021language} was used for this task. Results, calculated as Spearman's $\rho$ for the effect measured in each study's set of mutated sequences versus the predicted likelihood ratios over the WT counterparts, are shown in Table \ref{tab:dmsmutmeasurement} (Table \ref{tab:dmsmodelsystemmeasurement} shows the same results when aggregated to the organism level). Distributions of sequence lengths are further described for each of these modality groups in Appendix Tables~\ref{tab:dmslenstatsmeasure}, \ref{tab:dmslenstatsmodel}.  

\ExplSyntaxOn
\NewDocumentCommand { \getmax } { }
  {
    \clist_gset:Nx \g_tmpa_clist {\A, \B, \C, \E, \F}
    \clist_sort:Nn \g_tmpa_clist
      {
        \fp_compare:nNnTF {##1} < {##2}
          { \sort_return_swapped: }
          { \sort_return_same: }
      }
    \tl_gset:Nx \g_tmpa_tl { \clist_item:Nn \g_tmpa_clist {1} }
  }
% \NewDocumentCommand { \mynum } { m }
%   {
%     \fp_compare:nNnTF { #1 } = { \g_tmpa_tl }
%       { \cellcolor{blue!20} \num{#1} }
%       { \num{#1} }
%   }
\ExplSyntaxOff

\def\mywidth{1.7cm}
%\newgeometry{left=1.5cm,bottom=0.1cm}
\begin{table*}%[h]
  \sisetup{round-mode=places, round-precision=4}
  \centering
  \caption{\textbf{Zero-shot mutational effect analysis (experimental measurement modality).} Average Spearman's $\rho$ is shown, aggregated on the measurement modality of the DMS experiments. Number of studies for each modality shown in brackets. 'Winning' model for each modality is highlighted. }
  \begin{filecontents*}{sections/resultfiles/dmsmeasurement.csv}
,A,B,C,D,E,F,G
E1 reactivity (1),0.183895,0.212395,0.287169,0.27833,0.333237,0.3034,0.275613
Enzyme function (3),0.429029,0.437507,0.425796,0.430113,0.439879,0.430683,0.43255
Growth (20),0.441031,0.419773,0.446513,0.445785,0.458923,0.457495,0.454316
Ligase activity (1),0.416948,0.41532,0.429433,0.409679,0.395535,0.440491,0.440834
MIC (1),0.660696,0.637414,0.660246,0.655145,0.67029,0.667197,0.667917
Peptide binding (2),0.547703,0.493169,0.614554,0.617402,0.571913,0.585679,0.574061
Viral replication (5),0.392025,0.407516,0.362448,0.364072,0.385666,0.395059,0.372298
Yeast growth (3),0.395138,0.392979,0.394797,0.39957,0.404941,0.403025,0.40336
\end{filecontents*}
\csvreader[
  tabular=lp{\mywidth}p{\mywidth}p{\mywidth}p{\mywidth}p{\mywidth}p{\mywidth}p{\mywidth}p{\mywidth}p{\mywidth},
            table head=\toprule Measurement & \thead{ESM1b} & \thead{Pept.\\BMLM} & \thead{Pept.\\C-NPP}  & \thead{Pept.\\MLM} & \thead{Pept.\\NPP} \\ \midrule,
            head to column names,
            before line=\getmax,
            late after line = \\\hline,
            % late after last line=\\\bottomrule
            ]
            {sections/resultfiles/dmsmeasurement.csv}%
            {}%
            {\csvcoli & \mynum{\A} & \mynum{\B} & \mynum{\C} & \mynum{\E} & \mynum{\F} 
            }
  
  \label{tab:dmsmutmeasurement}
\end{table*}
%\restoregeometry
%}

%\end{document}

\paragraph{Evolutionary information enrichment in sequence embeddings.---}
Previous studies \cite{esm, prottrans, detlefsen2022learning} have illustrated that the representation space learned by protein models reflect evolutionary information. As such, without further supervision, these models are able to cluster evolutionarily related protein sequences closer than less related ones. We probe the impact of the continued training of protein models on chopped sequences on this ability. We sample common Pfam protein families~\cite{bateman2004pfam}, including beta-lactamase (PF00144), SH3 domain (PF00018), and WW domain (PF00397), and 
% Next, we reduce the representation spaces of the language models using t-SNE\cite{van2008visualizing} for a given Pfam family (Fig~\ref{fig:pfam}). 
apply t-SNE\cite{van2008visualizing}, for each family (Fig~\ref{fig:pfam}), on the extracted protein representations. 
% To evaluate the correlation between evolutionary information and the reduced representation space, 
We follow 
% the approach used in this study\cite{van2009dimensionality}: we
van der Maaten et al.\cite{van2009dimensionality} and 
calculate the generalization error from 1-nearest neighbor classifiers that are trained on the low-dimensional data representation. 
Our results (Table~\ref{tab:pfamclusters}) show that the BMLM peptide model outperforms ESM1b on beta-lactamase and WW families, while achieving competitive performance on SH3 family. Notably, the three representative protein families cover a range of sequence lengths: beta-lactamase (324 AAs), SH3 (47 AAs), and WW (30 AAs).

\ExplSyntaxOn
\NewDocumentCommand { \getminpfam } { }
  {
    \clist_gset:Nx \g_tmpa_clist {\untrained, \esmb, \mlm, \mlmnsp, \contrastive, \bmlm}
    \clist_sort:Nn \g_tmpa_clist
      {
        \fp_compare:nNnTF {##1} > {##2}
          { \sort_return_swapped: }
          { \sort_return_same: }
      }
    \tl_gset:Nx \g_tmpa_tl { \clist_item:Nn \g_tmpa_clist {1} }
  }
% \NewDocumentCommand { \mynum } { m }
%   {
%     \fp_compare:nNnTF { #1 } = { \g_tmpa_tl }
%       { \cellcolor{blue!20} \num{#1} }
%       { \num{#1} }
%   }
\ExplSyntaxOff

\def\mywidth{1.7cm}

\noindent\begin{table*}
  \sisetup{round-mode=places, round-precision=5}
  \centering
  \caption{\textbf{Clustering efficiency in recapitulating phyla memberships of proteins in three Pfam families.} For each family, performance is reported as average 1 nearest-neighbor generalization error for recovering phyla affiliations. Lower is better, lowest value for each Pfam family is shown in blue. }
  \begin{filecontents*}{sections/resultfiles/Pfam_clustering_errors_no_blosim_no_std.csv}
,untrained,esmb,mlm,mlmnsp,contrastive,bmlm
Beta-lactamase,0.2504,0.09355,0.09346,0.0975,0.12381,0.09333
SH3,0.33647,0.16854,0.16941,0.20633,0.23635,0.17371
WW,0.37625,0.2262,0.23574,0.28424,0.29571,0.22342
\end{filecontents*}
\csvreader[tabular=lp{\mywidth}p{\mywidth}p{\mywidth}p{\mywidth}p{\mywidth}p{\mywidth}p{\mywidth}p{\mywidth},
            table head=\toprule \thead{ Pfam family} & \thead{Untrained} & \thead{ESM1b} & \thead{Pept.\\MLM} & \thead{Pept.\\NPP} & \thead{Pept.\\C-NPP} & \thead{Pept.\\BMLM}\\ \midrule,
            head to column names,
            before line=\getminpfam,
            %late after line = \\\hline,
            late after last line=\\\bottomrule]
            {sections/resultfiles/Pfam_clustering_errors_no_blosim_no_std.csv}%
            {}%
            {\csvcoli & \mynum{\untrained} & \mynum{\esmb} & \mynum{\mlm} &
            \mynum{\mlmnsp} & \mynum{\contrastive} & \mynum{\bmlm}
            }
  
  \label{tab:pfamclusters}
\end{table*}

%\jg: long section header for this paragraph, will revisit. 
% \paragraph{Chopped sequences are distinct from natural peptides, and training only on chopped sequences results in more generalizable embeddings---}
\paragraph{Sequence type analysis - Chopped vs. Natural}

we compare our chopped-sequence models with models trained on natural peptides, derived either from short UR50-S sequences ($\sim1M$ peptides) or from Peptide Atlas~\cite{desiere2006peptideatlas} ($\sim3.5M$ sequences), which is the largest collection of mass-spectrometry identified peptides. In comparison to the chopped-sequence training, in which all $\sim30M$ UR50 sequences are considered, and the random on-the-fly chopping adds sequence variability between different epochs, we get a much larger data scale of almost $\sim30M * \#epochs$ (we use 25 epochs in our experimental setup, and assume here that most sequences are long enough to produce different unique chopped sequences at each epoch). We compare performance of each model across the various test sets and observe, at Table~\ref{tab:seq-type}, that while each model performs best on it's dedicated data type, the models trained on chopped sequences generalize better than others to unseen sequence types. 

\begin{figure}
    \centering
    \includegraphics[width=0.7\textwidth]{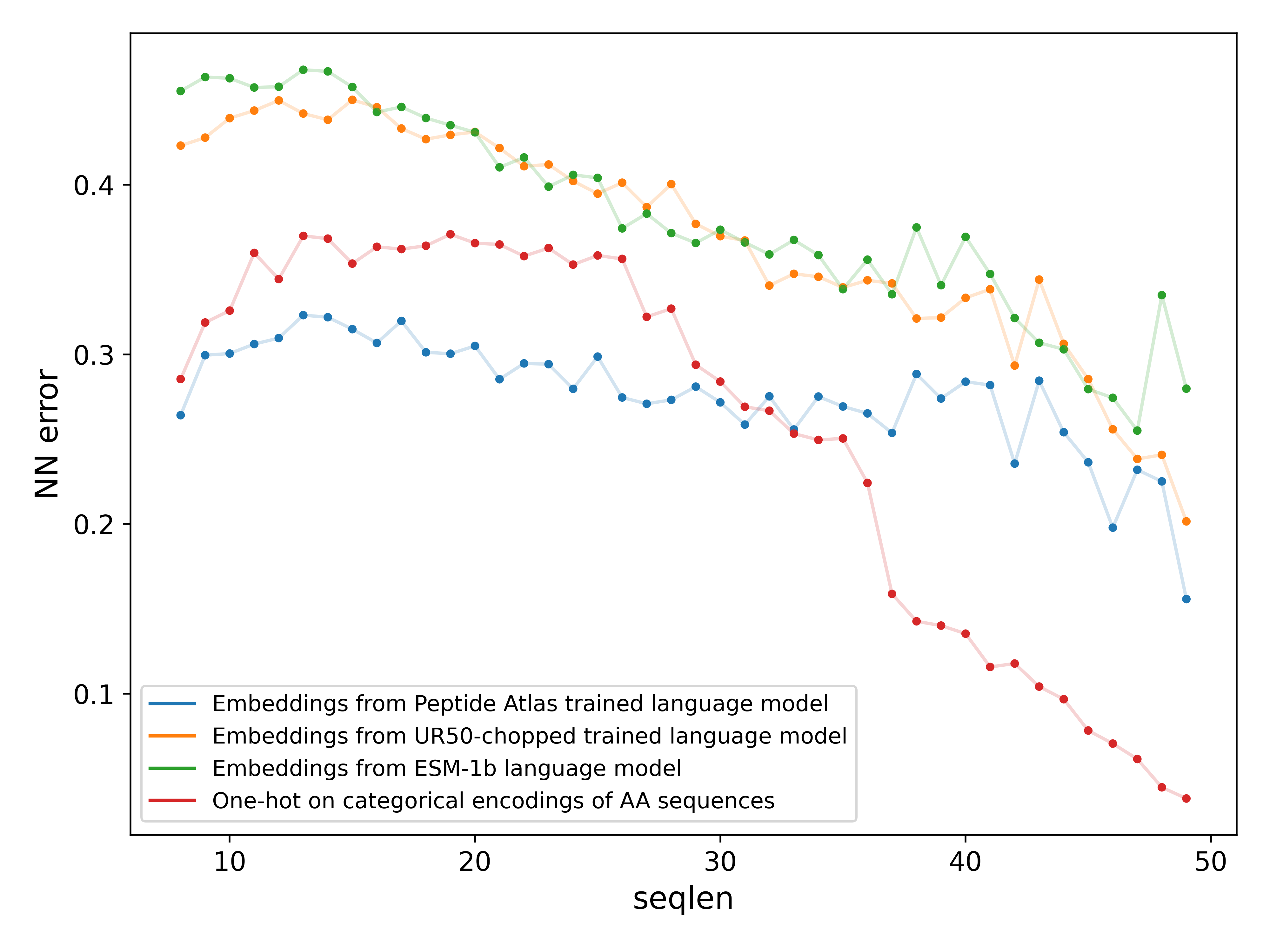}
    \caption{\textbf{Protein sequence representations encode evolutionary information.} The differences in peptide representations quantified by 1-NN generalization error on  test sequences is plotted at peptide lengths ranging from 8 to 50. Each line represents a distinct model. Lower error indicates the model embeddings from the two sequence populations (chopped and naturals) are more distinct. 
    % Higher error indicates the embeddings are similar.  
    }
    \label{fig:nnembed}
\end{figure}

Next, we seek to quantify the intrinsic differences of between chopped and natural peptide sequences and how such differences are perceived by different models (trained on different sequence types). We performed a balanced sampling of peptides from natural and chopped sequence test sets, and obtained their embeddings from models. To quantify the differences of those embeddings from the chopped and natural sequences, we examine the t-SNE projections of the embeddings and calculate the nearest-neighbor (NN) generalization error. Lower NN-error indicates the two populations are more distinctive whereas higher NN-error indicates the two populations are more intermixed in the embedding space. We use one-hot encoded sequence representations as the baseline trend for the intrinsic differences. ESM1b without finetuning was used as the baseline model. As shown in Fig~\ref{fig:nnembed}, both ESM1b and its chopped-sequence finetuned variant learn to bring the embeddings from two populations closer than the baseline (one-hot encoding), whereas language models finetuned on 
% a combination of chopped and naturals, or on 
naturals alone, learnt distinct embeddings for either set. All models showed a trend of generating more distinct embeddings for chopped and natural sequences as their length increases.  

% Additionally, we compare our chopped-sequence models with models trained on natural peptides, derived either from short UR50 sequences or from Peptide Atlas~\cite{desiere2006peptideatlas}, which is the largest collection of mass-spectrometry identified peptides. We compare performance of each model across the various test sets and observe, at Table~\ref{tab:seq-type}, that while each model performs best on it's dedicated data type, the models trained on chopped sequences (whether finetuned from ESM1b or trained from scratch) seem to generalize better than others to unseen sequence types. 

\ExplSyntaxOn
% \NewDocumentCommand { \getminece } { }
%   {
%     \clist_gset:Nx \g_tmpa_clist {\choppedft, \naturalft, \shortft, \chopped,\natural, \short}
%     \clist_sort:Nn \g_tmpa_clist
%       {
%         \fp_compare:nNnTF {##1} > {##2}
%           { \sort_return_swapped: }
%           { \sort_return_same: }
%       }
%     \tl_gset:Nx \g_tmpa_tl { \clist_item:Nn \g_tmpa_clist {1} }
%   }
% \NewDocumentCommand { \mynum } { m }
%   {
%     \fp_compare:nNnTF { #1 } = { \g_tmpa_tl }
%       { \cellcolor{blue!20} \num{#1} }
%       { \num{#1} }
%   }
\ExplSyntaxOff

\def\firstwidth{2.5cm}
\def\mywidth{1.5cm}

\noindent\begin{table*}[t]
%   \sisetup{round-mode=places, round-precision=4}
  \centering
  \caption{\textbf{Sequence type generalization analysis.} Comparison of models trained on natural vs. artificial ("chopped") short sequences. Models were trained with MLM objective. (FT) stands for models finetuned from ESM1b. We compare ECE (Lower is better) across test sets of each data type.}
%   \csvreader[
%             tabular=lP{\mywidth}P{\mywidth}P{\mywidth}P{\mywidth}P{\mywidth}P{\mywidth}P{\mywidth}P{\mywidth},
%             table head=\toprule \thead{Evaluation Set / Model-Type} & \thead{UR50-S \\ Chopped*} & \thead{Peptide \\ Atlas* } & \thead{UR50-S \\ Shorts*} & \thead{UR50-S \\ Chopped} & \thead{Peptide \\ Atlas} & \thead{UR50-S \\ Shorts}  \\ \midrule,
%             head to column names,
%             before line=\getminece,
%             % late after line = \\\hline,
%             late after last line=\\ \bottomrule]
%             {sections/resultfiles/data_type_results.csv}%
%             {}%
%             {\csvcoli & \mynum{\choppedft} & \mynum{\naturalft} & \mynum{\shortft} & \mynum{\chopped} & \mynum{\natural} & \mynum{\short}
%             }
    \begin{tabular}{ c | c | c c c | c c c}
    % {p{\firstwidth}p{\mywidth}p{\mywidth}p{\mywidth}p{\mywidth}p{\mywidth}p{\mywidth}}
    \hline
    \thead{Evaluation Set} & \thead{ESM1b} & \thead{UR50-S \\ Chopped (FT)} & \thead{Peptide \\ Atlas (FT) } & \thead{UR50-S \\ Shorts (FT)} & \thead{UR50-S \\ Chopped} & \thead{Peptide \\ Atlas} & \thead{UR50-S \\ Shorts} \\ \hline \hline
    UR50-S Chopped & 7.0837 & \textbf{5.9511} & 6.3835 & 6.3526 & 6.2687 & 7.6378 & 6.5923 \\ 
    Peptide Atlas & 7.4915 & 6.1710 & \textbf{5.3595} & 6.5857 & 6.3551 & 5.3868 & 6.7195\\ 
    UR50-S Shorts & 6.0321 & 5.7430 & 6.4114 & \textbf{5.6992} & 6.3257 & 8.3397 & 6.2595\\ \hline
    \end{tabular}            

  \label{tab:seq-type}
\end{table*}

%\jg : To add the new results from Gil for the models, and connect them to the previous section
%\jg : To-do adjust Figure 5 proportions and fix legend labels. 

\section{Conclusion}
Proteins perform many essential functions in biological systems and can be successfully developed as bio-therapeutics. 
% They are the emerging player after genomics in research and applications across academic and industrial space of healthcare and life sciences. Recently, many modern supervised and self-supervised ML methods have been developed to model the sequences, structures, and functions of various proteins. Most notably, DeepMind developed AlphaFold2~\cite{jumper2021highly}
% [Jumper et al. 2021 (https://www.nature.com/articles/s41586-021-03819-2)]
% , a protein structure prediction model that significantly improved performance over traditional approaches that model protein folding processes. These solutions can directly impact healthcare by accelerating drug discovery, drug design, and understanding the genetic basis of various diseases. 
%
%
%\hr{Not sure who is doing this -  [Something here about showing that focusing on a subset of the protein domain may be a preferred approach than training big subdomain-agnostic models in biology? Based on the pfam taxonomy embedding results, ECE of peptide transformers, linear probing on the various tasks]} 
%
Here we presented an approach for self-supervised training on shorter chopped protein sequences. Our empirical results showed that these learned representations are beneficial for downstream peptide-related tasks, while slightly sacrificing the performances on some of the longer protein-focused tasks.  Specifically, we observed peptide models outperformed ESM1b on context prediction accuracy in all length ranges, with improvement ranging from ~24.5\% on the longest to 41.7\% for the shortest sequence lengths.
We also found albeit having some differences with natural peptides, chopped protein sequences leads to models with improved generalization performance on out-of-distribution data. This indicates chopped proteins is a powerful data augmentation method for training protein language models.

\clearpage
\section*{Appendix}
\appendix
\label{section:appendix}
\subsection{Alternative training objectives}
\label{sec:appendix-training-objectives}

% \paragraph{Masked Language Modeling (MLM).---}
% This objective, originally proposed in BERT~\cite{devlin2018bert}, is commonly used for learning language and protein representations~\cite{esm, prottrans, heinzinger2019modeling, alley2019unified}. For each chopped protein sequence x, we sample a set of indices $M$ to mask, with probability p=0.15, replacing the true token at each index $i$ of the input sequence with either (1) the <MASK> token (p=0.8); (2) a random amino acid token (p=0.1); or (3) the unchanged $i^\mathrm{th}$ token (p=0.1). For each index $i \in M$, we independently minimize the negative log likelihood of the true amino acid $x_i$ given the masked input sequence $x_{/M}$ as context: $\mathcal{L}_{MLM} = \mathbb{E}_M[\sum_{i \in M} (\log{p(\hat{x}_i=x_i|x_{/M})}]$
% , where $\log{p(\hat{x}_i=x_i|x_{/M})}$ is the logit corresponding to the true $i^\mathrm{th}$ amino acid token. Figure~\ref{fig:pmlm} illustrates applying MLM on chopped proteins.

% \textbf{Next Peptide Prediction (NPP):}
\paragraph{Next Peptide Prediction (NPP).---}
For the NPP objective, firstly we reinterpret the `next sentence prediction' task from NLP, as proposed in the BERT paper~\cite{devlin2018bert}, as `next peptide prediction'. As in Ref.~\cite{devlin2018bert}, we concatenate two peptide sequences from the chopped protein dataset, with a special <SEP> token in between, as the input sequence to the model.  Given the first peptide before the <SEP> token, the second peptide is either the correct next peptide in the parent protein sequence or a randomly sampled one. The objective is a binary cross-entropy loss applied over a linear projection from the <CLS> token output representation. 
Despite the 'next-sentence-prediction' objective being criticized in recent NLP publications~\cite{xu2020symmetric, aroca2020losses}, no analogous research has been done for amino-acid sequences. We believe that due to the nature of the task, vocabulary and underlying chemical and structural considerations, NPP might be more challenging and hence lead to improved learned representations.

\paragraph{Contrastive NPP (C-NPP).---}
We introduce a contrastive extension of the NPP objective, where the model needs to learn to detect the correct next peptide sequence from the parent protein out of a large pool of candidates. Specifically, a batch of $N$ peptide samples will be composed of $N/2$ `first peptides' and their corresponding $N/2$ `next peptides'. 
The <CLS> token output representation is projected to a lower dimension, using two separate sets of learned projection weights. Cosine similarity is computed to provide an estimated compatibility score between each possible pair of  `first' vs `next' peptide sequences. 
The objective is composed of cross-entropy loss, following the contrastive objective defined in SimCLR~\cite{chen2020simple}, defined as follows for each positive pair of first and next peptides:
 $\mathcal    L_{\mathrm{C-NPP}}(i) = -\log{ \frac{\exp(S(f_i, n_i) / \tau)}{\sum_{k=0}^{N/2}{\exp(S(f_i, n_k)/\tau)}}}$,
where $f_i$ is the projected representation of the $i\mathrm{th}$ first peptide, $n_i$ is the projected representation of the corresponding $i\mathrm{th}$ next peptide, $S$ stands for cosine similarity between the two elements, and $\tau \in \mathbb R^+$ corresponds to a real temperature scaling factor. For efficient evaluation, the potential candidate pool is restricted to the true next peptides from different samples within the same batch. In addition, we extended this objective with an equivalent loss term for predicting the previous (first) peptide, from the pool of first peptide candidates, for each next peptide. MLM loss is applied in addition to these two loss terms. A similar contrastive objective has been shown to improve performance of protein language-style models trained on full protein sequences~\cite{shen2021improving}.

\begin{figure}[b]
    \centering
    \subcaptionbox{\label{fig:pmlm}}
    {
    \includegraphics[width=0.49\textwidth]{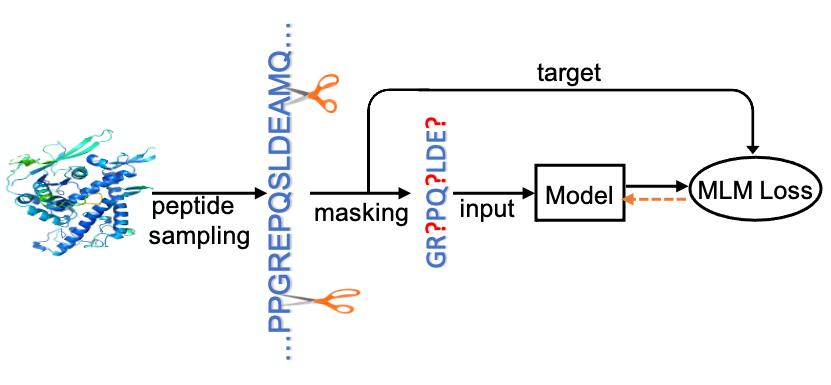}
    }
    \subcaptionbox{\label{fig:nsp}}
    {\includegraphics[width=0.24\textwidth,height=4cm]
    {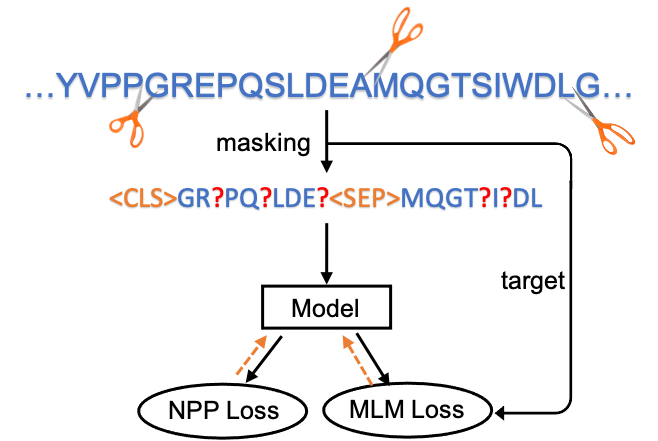}}
    % \hspace{1cm}
    \subcaptionbox{\label{fig:cnsp}}
    {\includegraphics[width=0.24\textwidth,height=4cm]
    {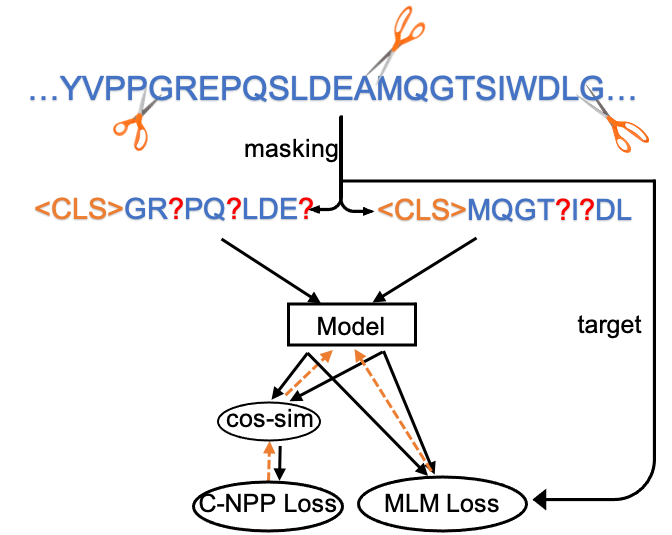}}
    \caption{\textbf{Self-supervised training procedures on chopped proteins.} 
    (a) Masked language modeling on chopped proteins.
    (b) NPP - a pair of chopped sequences is concatenated and processed jointly. A binary cross-entropy loss, over projected <CLS> token representation, optimizes classification of true vs fake next peptide.
    (c) C-NPP - first and next chopped peptides are processed separately. Compatibility between all first and next candidate peptides in the batch is computed via cosine similarity over projected <CLS> token representations. A contrastive cross-entropy loss is optimized to ensure correct pairs will receive higher compatibility scores than non-matching pairs.
    }
    \label{fig:NspLM}
\end{figure}

\paragraph{BLOSUM MLM (BMLM).---}
For the BLOSUM MLM objective, as in MLM, each input sequence is modified by replacing a fraction of the amino acid tokens with a special mask token. The network is trained to predict the missing tokens from the modified sequence. The idea is to "smooth" the one-hot removed token targets, based on the BLOSUM62 substitution matrix, and assign pseudo-probabilities to each amino-acid token based on the probability that it could substitute the removed token. The final objective is a KL-divergence loss between target and prediction distribution as follows: $\mathcal{L}_{BMLM} = \mathbb{E}_M[\sum_{i \in M} D_{KL}(p(\hat{x}_i|x_{/M}) || blosum({x}_i))]$, where $M$ is the set of masked elements, $D_{KL}$ denotes KL-divergence and $blosum(x_i)$ is the “pseudo-probability” function for the amino acid $x_i$. It returns a 1-d vector with 20 elements corresponding to a row in the BLOSUM matrix. The BLOSUM rows are transformed into pseudo-probabilities with Softmax.

\begin{figure}[t]
    \centering
    \includegraphics[width=0.5\textwidth]{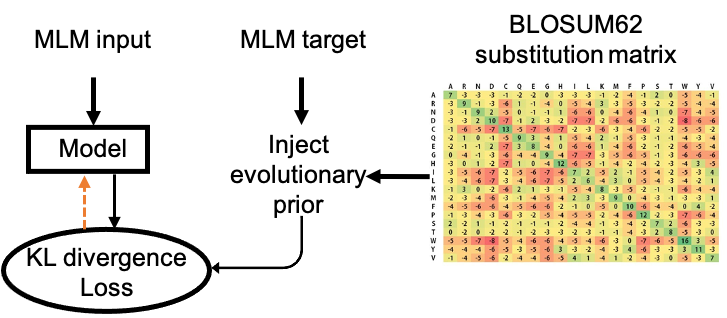}       
    % \includegraphics[width=0.8\textwidth]{sections/methodfigures/BLOSUM.png}       
    % \caption{\textbf{Training procedures incorporating evolutionary information utilizing the BLOSUM62 substitiution matrix.} \scriptsize{In BMLM (left), based on BLOSUM substitution scores of the target amino-acid, we transform the one-hot MLM target into a probability distribution over tokens, and minimize the KL-divergence loss. In BloSim (right), we randomly mutate a peptide, and compute cosine similarity between original and mutated sequence representations. We minimize MSE loss between predicted similarities and pre-computed BLOSUM similarity targets.
    % % produce peptide representations for which cosine similarity between them will yield similarity scores close to the pre-computed blosum similarity target scores.
    % In addition, we optimize a token-level binary cross-entropy loss over mutated sequences trying to detect the mutated tokens.}}
    % \label{fig:BlosumLM}
    \caption{\textbf{Training procedure incorporating evolutionary information utilizing the BLOSUM62 substitiution matrix}. In BLOSUM MLM (BMLM), we transform the one-hot MLM target into a probability distribution over tokens, based on the BLOSUM62 substitution scores of the target amino-acid, and minimize the KL-divergence loss.}
    \label{fig:BMLM}
\end{figure}

\begin{figure}[b]
    \centering
    \includegraphics[width=1\textwidth]{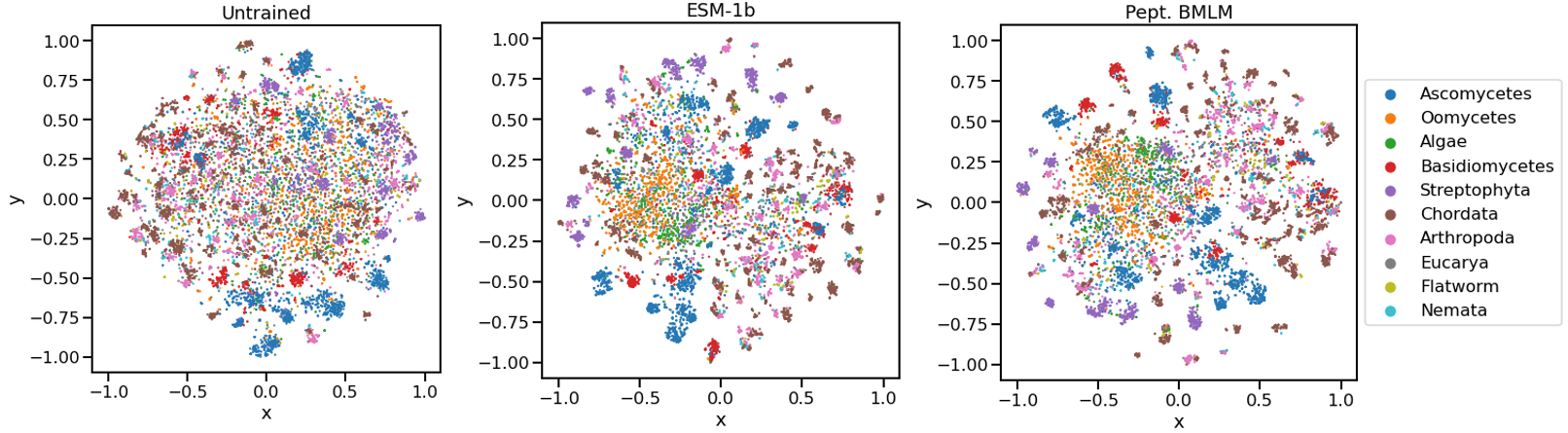}
    \caption{\textbf{Protein sequence representations encode evolutionary information.} Each point represents a protein/peptide sequence from WW domain (PF00397), and each sequence is colored by the phylum it belongs to (dimensionality is reduced by t-SNE). The panels correspond to sequence representations from untrained (left), ESM1b (middle), and peptide transformer with BMLM objective.}
    \label{fig:pfam}
\end{figure}

\subsection{Additional results}
\paragraph{Supervised evaluation on TAPE Stability task.---}
We evaluated the performance of protein and peptide language models on the TAPE stability prediction task~\cite{rao2019evaluating}. This task was selected because it contains exclusively peptides of length 45 amino acids. 
For this task, models were evaluated over 3 random seeds.
Training was done with early stopping until validation loss plateaued over 10 epochs. A warm-up schedule was applied for the learning rate with 5000 warm-up steps.

% For evaluation purposed, two versions of the TAPE secondary structure test datasets were generated, a) peptide subset with sequence lengths <= 50 amino-acids, and b) protein subset with sequence lengths >=50. 

Comparison of fine-tuning and linear probing on the TAPE stability task indicated improved generalization with linear probing. Results reported in this paper are from linear probing.   
%Specific parameters used for each task are outlined in the following table: 
% TAPE stability dataset contains exclusively peptides of length 45 amino acids. 
Two peptide transformers, namely those trained with MLM and BMLM objectives, significantly outperformed the ESM1b baseline protein language model on the stability task.  ESM1b is the current state-of-the art on this task, achieving a Spearman's $\rho$ of 0.7365.  When averaging over 3 random seeds, the Peptide BMLM and MLM models yield $\rho=0.7650 \pm 0.0067$ and $\rho=0.7673 \pm 0.0154$, respectively.  The other peptide objectives did not exceed the performance of ESM1b, giving $\rho=0.6707 \pm 0.0185$, and $\rho=0.6460 \pm 0.0141$, using C-NPP, and NPP, respectively.
%(Table \ref{tab:tapestability}).
To our knowledge, BMLM and MLM peptide transformers achieve a new state-of-the-art performance on this task.  

% \input{sections/resultfiles/TableTAPEstability.tex}
% The overall the performance of ESM-1b model closely matched that of peptide language models on the TAPE secondary structure task ('SS3'). In general, compared to protein test sequences, there was a performance drop across all models for the peptide test sequences in this dataset. However, the peptide language model with BLOSIM objective had a slight improvement over ESM-1b for the peptide subsets (0.8134 and 0.8120 accuracy respectively). Variance of the model performances across random seeds was minimal (<.00101 on average for proteins, <0.00842 on average for peptides).  
%\input{sections/resultfiles/TableTAPEss3proteins.tex}
%\input{sections/resultfiles/TableTAPEss3peptides.tex}

%\hl{Table or Figure 2. Performance of peptide language models with various objectives on downstream tasks. Panels A, B show evaluations on the TAPE tasks for protein stability and secondary structure prediction respectively. We observe improvement in performance over baseline ESM-1b when using the MLM objective with BLOSUM prior (‘bmlm’). }

% \hr{The paragraph aboves are very confusing to me. We don't have results for Sec. structure. @ZW or @Gs can you folks just compress this information down to a) stability task only and forget about the 5 TAPE task}

\setcounter{table}{0}
\renewcommand{\thetable}{A\arabic{table}}
%\lp{Can we expand this appendix to include the length distribution of all datasets we use?}
% \jg{To-do : Collate the length distributions into a table (We won't have this in time for AMLC)} 
\begin{comment}
\begin{figure}[h!]
  \caption{The spread of standard deviation ranges for each language model, across 3 random seeds (3 iterations per seed) is shown, as measured again accuracy calculations on the corresponding test sets in TAPE secondary structure prediction task. }
  \includegraphics[scale=0.25]{sections/resultfigures/std_ss3.png}
  \label{fig:ss3std}
\end{figure}
\end{comment}
\def\mywidth{1.7cm}
\begin{table*}[hp]
  \sisetup{round-mode=places, round-precision=1}
  \centering
  \caption{\textbf{Sequence length statistics for our UR50-S dataset splits.} Number of protein sequences shown in brackets in the first column.}
  \begin{filecontents*}{sections/resultfiles/statsUR50Seqlens.csv}
Split,A,B,C
Train (24.4M),311.7,16,36991
Validation (2.7M),311.6,16,34984
Test (3M),311.6,16,34674
\end{filecontents*}
\csvreader[tabular=lp{\mywidth}p{\mywidth}p{\mywidth},
            table head=\toprule Split & \thead{Average length} & \thead{Minimum length} & \thead{Maximum length} \\ \midrule,
            head to column names,
            %before line=\getmax,
            %late after line = \\\hline,
            %late after last line=\\\bottomrule
            ]
            {sections/resultfiles/statsUR50Seqlens.csv}%
            {}%
            {\csvcoli & \mynum{\A} & \mynum{\B} & \mynum{\C}
            }
  \label{tab:ur50lenstats}
\end{table*}
\def\mywidth{1.7cm}
\begin{table*}[hp]
  \sisetup{round-mode=places, round-precision=1}
  \centering
  \caption{\textbf{Sequence length statistics for DMS datasets, based on experimental measurement modality.} Number of studies for each modality shown in brackets in the first column.}
  \begin{filecontents*}{sections/resultfiles/statsDMSMeasurementSeqlens.csv}
Measurement,A,B,C
E1 reactivity (1),76,76,76
Enzyme function (3),345.333333333333,189,501
Growth (20),220.5,72,439
Ligase activity (1),104,104,104
MIC (1),263,263,263
Peptide binding (2),68.5,36,101
Viral replication (5),456.4,114,686
Yeast growth (3),164.333333333333,101,243
\end{filecontents*}
\csvreader[tabular=lp{\mywidth}p{\mywidth}p{\mywidth},
            table head=\toprule Measurement & \thead{Average length} & \thead{Minimum length} & \thead{Maximum length} \\ \midrule,
            head to column names,
            %before line=\getmax,
            %late after line = \\\hline,
            %late after last line=\\\bottomrule
            ]
            {sections/resultfiles/statsDMSMeasurementSeqlens.csv}%
            {}%
            {\csvcoli & \mynum{\A} & \mynum{\B} & \mynum{\C}
            }
  \label{tab:dmslenstatsmeasure}
\end{table*}

\begin{table*}[hp]
  \sisetup{round-mode=places, round-precision=1}
  \centering
  \caption{\textbf{Sequence length statistics for DMS datasets, based on model system for experiments.} Number of studies for each modality shown in brackets in the first column.}
  \begin{filecontents*}{sections/resultfiles/statsDMSModelssytemSeqlens.csv}
Organism,A,B,C
ECOLX (4),215.25,72,263
env (2),686,686,686
HUMAN (8),178.125,36,360
Other (12),331.916666666667,114,565
RODENT* (2),102.5,101,104
YEAST (8),119.375,75,240
\end{filecontents*}
\csvreader[tabular=lp{\mywidth}p{\mywidth}p{\mywidth},
            table head=\toprule Model system & \thead{Average length} & \thead{Minimum length} & \thead{Maximum length} \\ \midrule,
            head to column names,
            %before line=\getmax,
            %late after line = \\\hline,
            %late after last line=\\\bottomrule
            ]
            {sections/resultfiles/statsDMSModelssytemSeqlens.csv}%
            {}%
            {\csvcoli & \mynum{\A} & \mynum{\B} & \mynum{\C}
            }
  
  \label{tab:dmslenstatsmodel}
\end{table*}
\ExplSyntaxOn
% \NewDocumentCommand { \getmax } { }
%   {
%     \clist_gset:Nx \g_tmpa_clist {\A, \B, \C, \E, \F}
%     \clist_sort:Nn \g_tmpa_clist
%       {
%         \fp_compare:nNnTF {##1} < {##2}
%           { \sort_return_swapped: }
%           { \sort_return_same: }
%       }
%     \tl_gset:Nx \g_tmpa_tl { \clist_item:Nn \g_tmpa_clist {1} }
%   }
% \NewDocumentCommand { \mynum } { m }
%   {
%     \fp_compare:nNnTF { #1 } = { \g_tmpa_tl }
%       { \cellcolor{blue!20} \num{#1} }
%       { \num{#1} }
%   }
\ExplSyntaxOff

\def\mywidth{1.7cm}
%\newgeometry{left=1cm,bottom=0.1cm}
\noindent\begin{table*}[!htbp] 
  \sisetup{round-mode=places, round-precision=4}
  \centering
  \caption{\textbf{Zero-shot mutational effect analysis (experimental model system).} Average Spearman's $\rho$ is shown, aggregated on the model system used in the DMS experiments. Number of studies for each modality shown in brackets in the first column. 'Winning' model for each modality is highlighted.}
  \begin{filecontents*}{sections/resultfiles/dmsModelsystem.csv}
,A,B,C,D,E,F,G
HUMAN (8),0.394788,0.377776,0.395407,0.390239,0.393275,0.390712,0.389865
ECOLX (4),0.435808,0.412143,0.42453,0.417169,0.430301,0.428665,0.432899
env (2),0.465527,0.433645,0.368087,0.386802,0.466792,0.444149,0.442065
RODENT* (2),0.417434,0.389566,0.478852,0.473933,0.451181,0.471506,0.462674
Other (12),0.45364,0.455142,0.453579,0.452298,0.451954,0.459632,0.449258
YEAST (8),0.437382,0.424632,0.476503,0.482931,0.507166,0.502689,0.492117
\end{filecontents*}
\csvreader[tabular=lp{\mywidth}p{\mywidth}p{\mywidth}p{\mywidth}p{\mywidth}p{\mywidth}p{\mywidth}p{\mywidth}p{\mywidth},
            table head=\toprule Model system & \thead{ESM1b} & \thead{Pept.\\BMLM} & \thead{Pept.\\C-NPP}  & \thead{Pept.\\MLM} & \thead{Pept.\\NPP} \\ \midrule,
            head to column names,
            before line=\getmax,
            %late after line = \\\hline,
            late after last line=\\\bottomrule]
            {sections/resultfiles/dmsModelsystem.csv}%
            {}%
            {\csvcoli & \mynum{\A} & \mynum{\B} &%
             \mynum{\C} & \mynum{\E} & \mynum{\F}
            }
  \label{tab:dmsmodelsystemmeasurement}
\end{table*}
%\restoregeometry

\clearpage
\bibliographystyle{plain}
\bibliography{main}

\end{document}